\documentclass[aps,prl,twocolumn,superscriptaddress,nofootinbib]{revtex4-2}

\usepackage[T1]{fontenc}
\usepackage[utf8]{inputenc}
\usepackage{amsmath,amssymb,amsfonts,mathtools}
\usepackage{bm}
\usepackage{physics}
\usepackage{hyperref}
\usepackage{xcolor}
\usepackage{microtype}

\hypersetup{
  colorlinks=true,
  linkcolor=blue!60!black,
  citecolor=blue!60!black,
  urlcolor=blue!60!black
}

\newcommand{\SM}{\mathrm{SM}}

\newcommand{\Lag}{\mathcal{L}}

\newcommand{\Ocal}{\mathcal{O}}
\newcommand{\PL}{P_L}
\newcommand{\PR}{P_R}

\begin{document}

\title{Can a Nonstandard Invisible Pair Mimic the Michel Distribution?}

\author{Pablo Roig}
\email{pablo.roig@cinvestav.mx}
\affiliation{Departamento de Física, Cinvestav, IPN, Ciudad de México, Mexico}

\begin{abstract}
We ask whether a measured Michel distribution, apparently in excellent agreement with the Standard Model interpretation of the
\(
\ell_i \to \ell_j \nu\bar\nu
\) decay,
could instead arise from a different invisible sector.
Within a general low-energy effective field theory, we analyze lepton decays
\(
\ell_i \to \ell_j + X\bar X
\)
for electrically neutral, color-singlet, 
mutually conjugate invisible pairs \(X\bar X\) of spin up to $2$,
allowing (pseudo)scalar, (axial)vector, and antisymmetric tensor interactions in the lepton current, focusing on the massless limit relevant for exact degeneracies.
We formulate a criterion for indistinguishability based on the full set of measurable differential distributions. Under these assumptions, besides the obvious spin $1/2$ case, there is a unique nontrivial solution: a massless complex scalar pair coupled through a purely left-handed vector current exactly reproduces the standard Michel pattern, including its extensions to daughter-lepton polarization and radiative channels. 
All other cases studied here are distinguishable, in particular because higher-spin invisible pairs produce additional kinematic prefactors. 
These results isolate the only nonstandard and nontrivial invisible sector that can remain hidden in Michel-type lepton decay measurements. Phenomenologically, a subpercent branching fraction into this channel could bias the conventional muon-lifetime extraction of $G_F$ upward by enough for the corrected value to agree with the CKM-unitarity determination, while increasing the discrepancy with the electroweak fit.
\end{abstract}

\maketitle

Leptonic muon and tau decays provide some of the cleanest tests of the Lorentz structure of the charged weak current. In the standard treatment, the energy-angle 
distribution of the final charged lepton is parametrized by the Michel parameters
\(
(\rho,\eta,\xi,\delta)
\),
which take the values
\(
\rho=\delta=3/4
\),
\(
\xi=1
\),
and
\(
\eta=0
\)
in the Standard Model (SM)~\cite{Michel:1949qe,Bouchiat:1957zz,Kinoshita:1957zza,Fetscher:1986uj,Fetscher:1990su,ParticleDataGroup:2024cfk}.

The same framework extends to daughter-lepton polarization, while radiative decays with a real or pair-converted photon are described by generalized Michel parameters and richer differential distributions~\cite{Flores-Tlalpa:2015vga,Arbuzov:2016ywn}.
At current precision, QED radiative corrections are indispensable in extracting Michel parameters from muon decay data~\cite{Behrends:1955mb,Kinoshita:1957zz,Arbuzov:2002cn,Arbuzov:2004wr,Anastasiou:2005pn,Fael:2015gua,Arbuzov:2023qgc,Voznaya:2024jzl}. The most accurate measurements in muon decay yield~\cite{Pich:2013lsa,Gorringe:2015cma}~\footnote{The result for $\xi$ uses that the muon, from pion decay, is fully longitudinally polarized in very good approximation, with 
$P_\mu>0.99682$ at $90\%$ confidence level (C.L.)~\cite{Fetscher:1984da,Jodidio:1986mz,Langacker:1988fp}. Otherwise, results in eq.~(\ref{eq_currentlimits}) still apply, with the substitution $\xi\to P_\mu\xi$.}
\begin{eqnarray}\label{eq_currentlimits}
\rho=0.74979\pm0.00026\,&&\quad \eta=0.057\pm0.034,\nonumber\\
\xi=1.0009^{+0.0016}_{-0.0007} \,&&\quad \xi\delta=0.7511^{+0.0012}_{-0.0006}\,.
\end{eqnarray}
Notably, assuming lepton universality, both lepton decay modes of the tau lepton allow to extract $\eta=0.013\pm0.020$, which is the current most precise limit, in good agreement with the SM prediction, as is also the case for $\rho$, $\xi$ and $\delta$ in (\ref{eq_currentlimits}).

 These parameters have also been used to constrain several classes of new-physics effects (e.g. \cite{Santamaria:1986kg,Pich:1995vj,Stahl:1996gu,Dova:1997fj,Heeck:2024uiz,Heeck:2025jfs}). In this work, we take a complementary perspective: rather than asking how new particles modify the Michel parameters, we identify which invisible final states can remain hidden within the already measured SM-like spectrum because their energy and angular distributions are degenerate with the SM one. This raises our central question: does an experimentally SM-like Michel distribution really imply that the invisible pair is \(\nu\bar\nu\)? The answer is subtle, because lepton decays only probe the invisible system inclusively (see, instead \cite{Breso-Pla:2025cul}).
Indeed, the Particle Data Group explicitly notes that when invisible final-state particles are massless and not directly observed, certain intrinsic properties of the invisible sector cannot be directly tested through these decays alone~\cite{ParticleDataGroup:2024cfk}.
Motivated by this observation, we study the most general low-energy effective field theory (LEFT) for
\begin{equation}
\ell_i^-(p,s)\to \ell_j^-(p') + X(k_1)+\bar X(k_2),
\label{eq:process}
\end{equation}
where \(X\bar X\) is a neutral, color-singlet, mutually conjugate, nearly massless invisible pair, and ask under which conditions the observable distributions associated to (\ref{eq:process}) can be exactly indistinguishable from those of the SM Michel process
. Related effective descriptions of lepton decays into invisible pairs were developed in Refs.~\cite{He:2022ljo,Liang:2023yta}. More recently, Ref.~\cite{Jahedi:2025hnu} considered dark-matter candidates using branching ratios, the invisible-pair invariant-mass spectrum, radiative decays and invisible muonium decays, up to spin one. Here we address the complementary question of which invisible sectors are exactly indistinguishable from the SM in Michel-type observable distributions.

We define \emph{indistinguishability} operationally: a nonstandard invisible pair \(X\bar X\) is indistinguishable from the SM if all differential distributions built exclusively from measurable degrees of freedom coincide with those of the SM up to an overall normalization.
This includes the standard Michel spectrum, the dependence on the initial-lepton polarization, the polarization of the daughter charged lepton, and the measurable distributions in the radiative channels.

We consider scalar, vector, and antisymmetric tensor interactions in the lepton current, allowing both charged-lepton chiralities.
The kinematics is controlled by the total invisible momentum (working in the decaying-lepton rest frame)
\begin{equation}
q^\mu \equiv k_1^\mu+k_2^\mu = p^\mu-p'^\mu,
\qquad
q^2 = m_{\ell_i}^2+m_{\ell_j}^2-2m_{\ell_i}E_{\ell_j},
\end{equation}
or equivalently by the reduced charged-lepton energy
\begin{equation}
x \equiv \frac{2E_{\ell_j}}{m_{\ell_i}}.
\end{equation}
For the SM Michel decay of $\bar{\ell_i}$ particles ($\theta$ is the angle between $\vec{p'}$ and the decaying particle polarization vector, $\vec{P}_{\ell_i}$, with $0\leq P_{\ell_i}=\big|\vec{P}_{\ell_i}\big|\leq 1$),
\begin{equation}
\frac{\mathrm{d}^2\Gamma_{\SM}}{\mathrm{d} x\,\mathrm{d}\cos\theta}
\propto
x^2\left[(3-2x)+P_{\ell_i}\cos\theta(2x-1)\right]
\qquad
(m_{\ell_j}\to0),
\label{eq:MichelSM}
\end{equation}
up to the well-known QED radiative deformations of the inclusive spectrum~\cite{Behrends:1955mb,Kinoshita:1957zz,Arbuzov:2002cn,Arbuzov:2004wr,Anastasiou:2005pn,Arbuzov:2023qgc,Voznaya:2024jzl}.

Our central result is simple: Within the class of single leading local structures considered here (see table \ref{tab:invisible_currents_shapes}), and in the massless invisible limit, the only nontrivial exact mimic of the SM Michel pattern is a massless complex scalar pair coupled through a purely left-handed vector current~\footnote{If lower-dimensional scalar-current operators are also present, see $\mathcal{O}^{(\phi)}_S$ in eq.~(\ref{eq:scalarops}), their coefficients must be sufficiently suppressed, since their Michel shapes are not SM-like, see table \ref{tab:invisible_currents_shapes}.},
\begin{equation}
\boxed{
\Lag_{\rm eff}\supset
\frac{C_\phi}{\Lambda^2}
\big(\phi^\dagger \overleftrightarrow{\partial_\mu}\phi\big)\,
\bar\ell_j\gamma^\mu \PL \ell_i
+\mathrm{h.c.},
\qquad
m_\phi=0.
}
\label{eq:mainoperator}
\end{equation}
In this case, the integrated unobservable invisible tensor is proportional to the same transverse projector as in the massless \(\nu\bar\nu\) case, so that the full measurable distribution predicted at tree level coincides with Eq.~\eqref{eq:MichelSM} up to normalization
. This degeneracy extends to daughter-lepton polarization and, as argued below, to the measurable structure of radiative and internally converted channels.

Before discussing other possible interactions and showing the uniqueness of this non-trivial degeneracy, let us briefly summarize the possible phenomenological consequences of this result.

As is well-known~\cite{ParticleDataGroup:2024cfk}, the Fermi constant, $G_F$, is one of the fundamental parameters of the Standard Model, which is fixed by muon decay. Less precisely, it can also be determined from the electroweak (EW) fit, or from first row CKM unitarity. In this case, an effective Fermi constant associated with that CKM-unitarity test can be defined as
$G_F^{\mathrm{CKM}} \equiv \sqrt{|V_{ud}|^2+|V_{us}|^2+|V_{ub}|^2}\, G_F^{\mu\text{-decay}}$,
where the numerically negligible $|V_{ub}|^2$ term will be omitted below. In the SM all three $G_F$ determinations should be consistent, while their possible discrepancy would hint at new physics, within its significance~\cite{Marciano:1999ih,Crivellin:2021njn}. The $G_F$ extraction from muon decay has just been improved~\cite{Eberhart:2026klz}, to
$G_F^{\mu\rm{-decay}}=1.16637859(59)\times10^{-5}$ GeV$^{-2}$. From the global EW fit, it is found instead~\cite{Crivellin:2021njn} $G_F^{EW\rm-fit}=1.16716(39)\times10^{-5}$ GeV$^{-2}$, which is $2.0\sigma$ above $G_F^{\mu\rm{-decay}}$. Finally, using $|V_{ud}|=0.97367(32)$ and $|V_{us}|=0.22431(85)$~\cite{ParticleDataGroup:2024cfk}, one gets $G_F^{CKM}=\sqrt{|V_{ud}|^2+|V_{us}|^2}\times G_F^{\mu\rm{-decay}}=1.16541(43)\times10^{-5}$ GeV$^{-2}$, which is $2.3\sigma$ below $G_F^{\mu\rm{-decay}}$.

In this context, the decay $\mu\to e \phi\bar\phi$ (for small enough $m_\phi$, see discussion below) would be indistinguishable from $\mu\to e\nu \bar\nu$, and would imply that $\Gamma_\mu=\tau_\mu^{-1}$ is not saturated by the SM Michel decay. Conventionally (conv.), $G_F^{\mu\text{-decay}}$ is extracted by identifying the measured total width with the SM prediction,

\begin{equation}
\Gamma_\mu^{\rm tot}\equiv\tau_\mu^{-1} \stackrel{\rm \tiny{conv.}}{=} \Gamma_{\rm SM}(\mu\to e\nu\bar\nu) = (G_F^{\mu\text{-decay}})^2 \frac{m_\mu^5}{192\pi^3}(1+\Delta q),
\end{equation}
with $\Delta q$ including higher-order corrections and finite electron-mass effects~\cite{Eberhart:2026klz}. This $G_F^{\mu\text{-decay}}$ would be larger than $G_F$, because of the non-vanishing branching ratio into the $e\phi\bar\phi$ decay channel.

Following ref.~\cite{Jahedi:2025hnu}, we will use as bound for $\mu\to e \phi \bar{\phi}$ the limit for the lepton flavor violating $\mu\to e \nu_e \bar{\nu}_\mu$ decay branching ratio~ ($<1.2\%$ at $90\%$ C.L.~\cite{ParticleDataGroup:2024cfk}). In this way, the upper limit $BR(\mu\to e\phi \bar{\phi})<1.2\%$ would allow an upward bias of up to approximately $0.6\%$ in the value of $G_F$ conventionally extracted from the muon lifetime. This is about seven times larger than the shift required to bring the underlying Fermi constant into agreement with the CKM-unitarity determination.

If the invisible branching fraction were such that the corrected muon-decay extraction coincided with the central value of $G_F^{\rm CKM}$, and if this branching fraction were known with negligible uncertainty, the resulting muon-decay determination would lie $4.5\sigma$ below $G_F^{\rm EW-fit}$. The direct comparison between the CKM and electroweak-fit determinations, retaining the present CKM uncertainty, corresponds instead to a $3.0\sigma$ difference.

\begin{figure}[t] \centering \includegraphics[width=0.92\columnwidth]{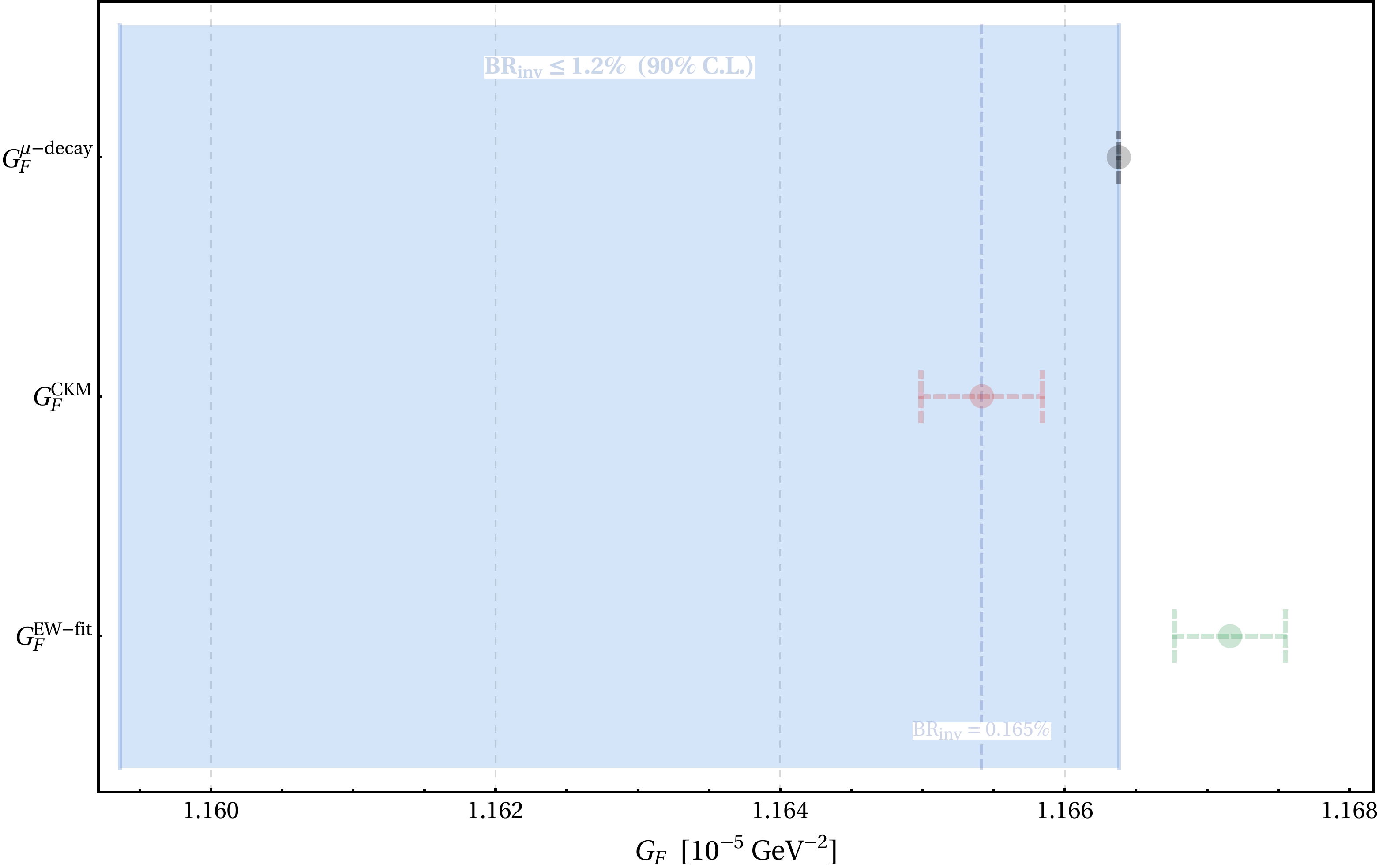} \caption{ Comparison of the Fermi constant extracted from the muon lifetime, from first-row CKM unitarity, and from the electroweak fit. The blue region shows the range of the underlying Fermi constant allowed by an invisible muon-decay branching fraction $\mathrm{BR}_{\rm inv}\leq 1.2\%$ at $90\%$ C.L. The dashed line marks the value of the underlying Fermi constant obtained for $\mathrm{BR}_{\rm inv}=0.165\%$, which coincides with the CKM determination.} \label{fig:GF_comparison} \end{figure}

$G_F^{EW\rm{-fit}}$ is pulled away from the others mainly by~\cite{Crivellin:2021njn} $M_W$, $\sin^2\theta_W$ from hadron colliders, the lepton asymmetry parameter $A_\ell$, and the forward-backward lepton asymmetry at the $Z^0$ peak, $A_{FB}^{0,\ell}$, observables at considerably higher energies than those determining $G_F^{\mu\rm{-decay}}$ and $G_F^{CKM}$, which makes $G_F^{EW\rm{-fit}}$ naturally more sensitive to heavy new physics contributions. The corresponding scale of this beyond-the-SM effect would be encoded in $\Lambda$ in eq.~(\ref{eq:mainoperator}). The constraint associated to producing $BR<1.2\%$ at $90\%$ C.L. is
 \begin{equation}
 \frac{|C|}{\Lambda^2}\leq 5.14\times10^{-6}\,\rm{GeV}^{-2}.
 \end{equation}
 For $C=1$ it would correspond to $\Lambda\gtrsim441$ GeV. Alternatively, if we demand that the shift in $G_F^{\mu\rm{-decay}}$ makes it match $G_F^{CKM}$, $BR(\mu\to e\phi\bar\phi)\sim0.165\%$ is required, corresponding, for $C=1$, to $\Lambda\simeq726$ GeV, which appears to be a plausible scale. Fig.~\ref{fig:GF_comparison} illustrates this discussion.

A natural question is then how light needs the $\phi$ particle be, so that its mass effects do not allow distinguishing it from the SM Michel decay? For $m_\phi\neq0$, the exact degeneracy is lifted both by the shift of the electron-energy endpoint,
\begin{equation}
\Delta E_e^{\max}=\frac{2m_\phi^2}{m_\mu},
\end{equation}
and by the massive two-body threshold factor
\begin{equation}
\left(1-\frac{4m_\phi^2}{q^2}\right)^{3/2}\,.
\end{equation}
The experimental mass sensitivity is therefore not fixed by kinematics alone, but depends on the branching fraction, momentum resolution, acceptance and systematic uncertainties of the spectral fit. Existing Michel-parameter analyses assume massless invisible particles and do not provide a direct limit on $m_\phi$ for the operator considered here. A dedicated recast of the measured electron momentum-angle distribution would be required. For reference, $m_\phi=(2,5,10)\,\mathrm{MeV}$ shifts the endpoint by approximately $(0.076,0.473,1.89)\,\mathrm{MeV}$, respectively. For orientation, the TWIST two-body search \cite{TWIST:2014ymv} treated $m_X\simeq13$ MeV as the lower edge of the mass region where the spectral-line sensitivity becomes comparatively robust. Equating the corresponding endpoint displacement to that of a scalar pair gives the purely kinematic estimate $m_\phi\simeq m_X/2\simeq6.5\ {\rm MeV}$. Since the two-body signal is monoenergetic whereas $\mu\to e\phi\bar\phi$ produces a continuous spectrum, this correspondence should only be regarded as indicative. A dedicated fit would be required to establish the actual sensitivity.

After highlighting the phenomenological interest of the $\mu\to e\phi\bar\phi$ decays as a way to bring into agreement the low-energy determinations of $G_F$, as well as discussing its interpretation and feasibility, it is instructive to summarize the alternatives to eq.~(\ref{eq:mainoperator}).
For a neutral complex scalar \(\phi\), the lowest-dimension operators (we always focus on these, for a given class, see table \ref{tab:invisible_currents_shapes}) may be organized as
\begin{align}
\Ocal_S^{(\phi)} &=
(\phi^\dagger\phi)\,[
\bar\ell_j (g_L^S \PL + g_R^S \PR)\ell_i+\mathrm{h.c.}],
\nonumber\\
\Ocal_V^{(\phi)} &=
(\phi^\dagger \overleftrightarrow{\partial_\mu}\phi)\,
[\bar\ell_j \gamma^\mu (g_L^V \PL + g_R^V \PR)\ell_i+\mathrm{h.c.}],
\nonumber\\
\Ocal_T^{(\phi)} &=
(\partial_{[\mu}\phi^\dagger\partial_{\nu]}\phi)\,
[\bar\ell_j \sigma^{\mu\nu}(g_L^T \PL + g_R^T \PR)\ell_i+\mathrm{h.c.}].
\label{eq:scalarops}
\end{align}
Among them, only the purely left-handed vector case in Eq.~\eqref{eq:mainoperator} reproduces the SM Michel pattern.
The scalar and antisymmetric-tensor operators generate clearly different spectra, see table \ref{tab:invisible_currents_shapes}.

For a neutral fermion pair \(f\bar f\), a \(V\!-\!A\) interaction of the same form as in the SM is trivially degenerate with the standard case.
Mass effects~\cite{Prezeau:2004md} can be organized in terms of generalized Michel parameters, and when projected onto the standard basis \((\rho,\eta,\xi,\delta)\) the leading contribution is \(\eta\)-like, whereas the full problem with massive invisible fermions requires generalized Michel parameters beyond the standard set~\cite{Marquez:2022bpg} (see also \cite{Shrock:1981wq}). The well-known distributions for $s=1/2$ fermions are reproduced~\cite{Kinoshita:1957zz} and collected in table~\ref{tab:invisible_currents_shapes}.

For this case of $s=1/2$ invisible particles, the usual Fierz ambiguity of muon decay remains~\cite{Fetscher:1986uj}. In particular, in the massless limit, the operators $(\bar{e}P_Rf)(\bar{f}P_L\mu)$ (in a 'charge-changing' form) and $(\bar{e}\gamma^\alpha P_L\mu)(\bar{f}\gamma_\alpha P_R f)$ (in the 'charge-retention' form), can lead to identical Michel spectra after normalization. In the SM neutrino case, this ambiguity is broken only by processes involving external neutrinos, such as inverse muon decay. For generic invisible fermions $f$, no such scattering measurement exists, so this degeneracy cannot be resolved by $\mu\to e+$ invisible data alone. This Fierz degeneracy is distinct from the one identified in this work: a pair of massless scalar invisibles produced by a vector current can reproduce the same inclusive invisible tensor as the SM neutrino pair. Nevertheless, it must always be taken into account. Finally, if $f$ carries lepton number, and neglecting effects of $\mathcal{O}(m_f)$, there is a one-to-one correspondence between lepton-number-conserving and violating transitions, as in the neutrino case~\cite{Langacker:1988cm}, again because the final-state neutral fermions are unobserved.

The situation changes qualitatively for invisible spin \(1\).
A consistent massless limit requires a gauge-invariant formulation~\cite{Ibarra:2021xyk} which, in this case, is automatic in terms of the field-strength tensor \(V_{\mu\nu}\) (describing a massless invisible vector particle), and not the bare Proca field.
Vector interaction then takes the form
\begin{equation}
\Ocal_V^{(V)} =
\big(V^\dagger_{\alpha\beta}\overleftrightarrow{\partial_\mu}V^{\alpha\beta}\big)\,
[\bar\ell_j \gamma^\mu (g_L \PL + g_R \PR)\ell_i+\mathrm{h.c.}]
\label{eq:vectorop}
\end{equation}
For a purely left-handed lepton current ($V-A$)
, the measurable distribution becomes
\begin{equation}
\frac{\mathrm{d}^2\Gamma_{V-A}^{V}}{\mathrm{d} x\,\mathrm{d}\cos\theta}
\propto
x^2(1-x)^2
\left[(3-2x)+P_{\ell_i}\cos\theta(2x-1)\right],
\label{eq:vectorMichel}
\end{equation}
which differs from the SM by the additional global \((1-x)^2\) factor.
Thus, even the vector case for massless complex spin-1 invisibles is not degenerate with the SM Michel distribution. The cases with scalar or tensor interactions in Dirac space present additional differences with respect to the SM scenario, as sketched in table \ref{tab:invisible_currents_shapes}.

A completely analogous conclusion holds for a massless complex spin-\(3/2\) invisible pair.
In that case, the LEFT must again be gauge invariant, now using the Rarita--Schwinger field strength \(\Psi_{\mu\nu}\), leading to the operator (only the one with left-handed $\ell_i$ is displayed, as it gives the closest distribution to the SM)
\begin{equation}
\Ocal_{V-A}^{(3/2)} =
(\bar\Psi_{\alpha\beta}\gamma^\mu \PL \Psi^{\alpha\beta})\,
(\bar\ell_j\gamma_\mu \PL \ell_i+\mathrm{h.c.})
\label{eq:rsop}
\end{equation}
The resulting spectrum is again, as a consequence of using the minimal gauge-invariant formulation, of the form
\begin{equation}
\frac{\mathrm{d}^2\Gamma_{V-A}^{(3/2)}}{\mathrm{d} x\,\mathrm{d}\cos\theta}
\propto
x^2(1-x)^2
\left[(3-2x)+P_{\ell_i}\cos\theta(2x-1)\right],
\label{eq:rsmichel}
\end{equation}
as in eq.~(\ref{eq:vectorMichel}), and is therefore distinguishable from the SM. Once more, additional differences arise for scalar or tensor currents, as displayed in table \ref{tab:invisible_currents_shapes}.

The massless complex spin-$2$ case is even more constrained.
Under the standard assumptions of locality, Lorentz invariance, and flat spacetime, Weinberg--Witten-type no-go theorems and related high-spin consistency arguments severely restrict the existence of complex massless higher-spin states coupled to Lorentz-covariant conserved currents~\cite{Weinberg:1964ew,Weinberg:1965rz,Weinberg:1980kq,Porrati:2008rm}.
For this reason, we do not regard the massless complex spin-2 case (see table \ref{tab:invisible_currents_shapes}) as a robust member of the same LEFT class.
Formally, if one forces a gauge-invariant construction
, 
even stronger kinematic prefactors appear, schematically of the form \((1-x)^{3,4}\), making this scenario manifestly distinguishable from the SM.

The three-body analysis is not the end of the story.
Radiative leptonic decays are described by generalized Michel parameters~\cite{Arbuzov:2016ywn}, while five-body channels
\(
\ell_i\to \ell_j\ell'^+\ell'^-+\text{invisibles}
\)
probe the full energy-angle dependence of three charged leptons~\cite{Flores-Tlalpa:2015vga}.
For the scalar degeneracy in Eq.~\eqref{eq:mainoperator}, the same LEFT reasoning indicates that neither the radiative channel nor the internally converted five-body process lifts the degeneracy, provided the invisible sector is massless and electrically neutral, so that only the observable lepton current radiates in QED.
The reason is that, after integrating over the massless invisible pair, the same transverse projector is obtained as in the \(\nu\bar\nu\) case.
QED radiative corrections are known to modify the functional form of the inclusive muon spectrum and must be included before fitting Michel parameters~\cite{Kinoshita:1957zz,Anastasiou:2005pn,Arbuzov:2023qgc,Voznaya:2024jzl}. 
However, for the scalar case in Eq.~\eqref{eq:mainoperator}, by identical argument, QED corrections act on the same left-handed charged-lepton current as in the SM, while the integrated invisible tensor remains unchanged, so that the same factorization argument implies that the QED-corrected inclusive spectrum remains degenerate up to normalization, within this effective description and neglecting effects that resolve the invisible sector.

In the limit considered here, the polarization-independent and polarization-dependent functions that define the Michel spectrum are polynomials in \(x\). Thus, a fine-tuned linear combination of several operators could, in principle, mimic the SM polynomial shape. Nevertheless, for the finite set of local structures displayed in table~\ref{tab:invisible_currents_shapes}, this possibility is absent once the boldface entries are excluded. Additional derivatives merely introduce extra powers of \(q^2/m_\mu^2=1-x\), or higher-partial-wave tensors, and do not generate a new SM-like degeneracy. Interference effects can survive in the chiral limit and modify the Michel functions, but they do not enlarge the set of SM-degenerate possibilities because of the required simultaneous proportionality of the isotropic and anisotropic functions to the SM shapes.

In summary, we have identified the unique nontrivial invisible sector that can remain hidden to Michel-type measurements alone.
Among all neutral, color-singlet, mutually conjugate nearly massless invisible pairs considered here, only a massless complex scalar pair with a purely left-handed vector coupling can exactly mimic the standard Michel distribution and its observable extensions.
Invisible pairs of spin \(1\) and \(3/2\) cannot be exactly degenerate with the SM shape, because they induce observable kinematic distortions even with a left-handed lepton current, as well as the massless complex spin-2 case (whose consistency is in any case questionable under the assumptions stated above). Our results are summarized in table \ref{tab:invisible_currents_shapes}. 
Thus, even a Michel spectrum in excellent agreement with the SM does not uniquely imply \(\nu\bar\nu\). On the contrary, it constrains nonstandard invisible sectors far more sharply than is often assumed.

\begin{table*}[t]
\centering
\renewcommand{\arraystretch}{1.45}
\setlength{\tabcolsep}{4.5pt}
\begin{ruledtabular}
\begin{tabular}{c c p{0.40\linewidth} c c}
$s_X$ &
Interaction type &
Invisible current $J_X$ &
$F_0(x)$ &
$F_P(x)$
\\
\hline
$0$ &
$S$ &
$\displaystyle J_S^{(0)}=\phi^\dagger\phi$
&
$\displaystyle x^2$
&
$\displaystyle x^2$
\\[1mm]

$0$ &
$V$ &
$\displaystyle J_V^{(0)\mu}
=i\,\phi^\dagger\overleftrightarrow{\partial^\mu}\phi$
&
$\displaystyle \mathbf{x^2(3-2x)}$
&
$\displaystyle \mathbf{-x^2(2x-1)}$
\\[1mm]

$0$ &
$T$ &
$\displaystyle J_T^{(0)\mu\nu}
=\partial^{[\mu}\phi^\dagger\,\partial^{\nu]}\phi$
&
$\displaystyle x^2(1-x)(3-x)$
&
$\displaystyle x^2(1-x)(1+x)$
\\[1mm]
\hline

$\tfrac12$ &
$S$ &
$\displaystyle J_{S(P)}^{(1/2)}=\bar f (i\gamma_5) f$
&
$\displaystyle x^2(1-x)$
&
$\displaystyle x^2(1-x)$
\\[1mm]

$\tfrac12$ &
$V$ &
$\displaystyle J_{V(A)}^{(1/2)\mu}=\bar f\gamma^\mu (\gamma_5) f$
&
$\displaystyle \mathbf{x^2(3-2x)}$
&
$\displaystyle \mathbf{-x^2(2x-1)}$
\\[1mm]

$\tfrac12$ &
$T$ &
$\displaystyle J_T^{(1/2)\mu\nu}=\bar f\sigma^{\mu\nu}f$
&
$\displaystyle x^2(3-x)$
&
$\displaystyle x^2(1+x)$
\\[1mm]
\hline

$1$ &
$S$ &
$\displaystyle J_S^{(1)}
=V^\dagger_{\alpha\beta}V^{\alpha\beta}$
&
$\displaystyle x^2(1-x)^2$
&
$\displaystyle x^2(1-x)^2$
\\[1mm]

$1$ &
$V$ &
$\displaystyle J_V^{(1)\mu}
=i\,V^\dagger_{\alpha\beta}
\overleftrightarrow{\partial^\mu}
V^{\alpha\beta}$
&
$\displaystyle x^2(1-x)^2(3-2x)$
&
$\displaystyle -x^2(1-x)^2(2x-1)$
\\[1mm]

$1$ &
$T$ &
$\displaystyle J_T^{(1)\mu\nu}
=i\left(
V^{\dagger\mu}{}_{\alpha}V^{\nu\alpha}
-
V^{\dagger\nu}{}_{\alpha}V^{\mu\alpha}
\right)$
&
$\displaystyle x^2(1-x)(3-x)$
&
$\displaystyle x^2(1-x)(1+x)$
\\[1mm]
\hline

$\tfrac32$ &
$S$ &
$\displaystyle J_S^{(3/2)}
=\bar\Psi_{\alpha\beta}\Psi^{\alpha\beta}$
&
$\displaystyle x^2(1-x)^3$
&
$\displaystyle x^2(1-x)^3$
\\[1mm]

$\tfrac32$ &
$V$ &
$\displaystyle J_V^{(3/2)\mu}
=\bar\Psi_{\alpha\beta}\gamma^\mu\Psi^{\alpha\beta}$
&
$\displaystyle x^2(1-x)^2(3-2x)$
&
$\displaystyle -x^2(1-x)^2(2x-1)$
\\[1mm]

$\tfrac32$ &
$T$ &
$\displaystyle J_T^{(3/2)\mu\nu}
=\bar\Psi_{\alpha\beta}\sigma^{\mu\nu}\Psi^{\alpha\beta}$
&
$\displaystyle x^2(1-x)^2(3-x)$
&
$\displaystyle -x^2(1-x)^2(1+x)$
\\[1mm]
\hline

$2$ (formal) &
$S$ &
$\displaystyle J_S^{(2)}
=R^\dagger_{\alpha\beta\rho\sigma}
R^{\alpha\beta\rho\sigma}$
&
$\displaystyle x^2(1-x)^4$
&
$\displaystyle x^2(1-x)^4$
\\[1mm]

$2$ (formal) &
$V$ &
$\displaystyle J_V^{(2)\mu}
=i\,R^\dagger_{\alpha\beta\rho\sigma}
\overleftrightarrow{\partial^\mu}
R^{\alpha\beta\rho\sigma}$
&
$\displaystyle x^2(1-x)^4(3-2x)$
&
$\displaystyle -x^2(1-x)^4(2x-1)$
\\[1mm]

$2$ (formal) &
$T$ &
$\displaystyle J_T^{(2)\mu\nu}
=i\left(
R^{\dagger\mu}{}_{\alpha\rho\sigma}
R^{\nu\alpha\rho\sigma}
-
R^{\dagger\nu}{}_{\alpha\rho\sigma}
R^{\mu\alpha\rho\sigma}
\right)$
&
$\displaystyle x^2(1-x)^3(3-x)$
&
$\displaystyle -x^2(1-x)^3(1+x)$
\end{tabular}
\end{ruledtabular}
\caption{
Representative leading invisible currents~\footnote{We recall that $A\overleftrightarrow{\partial^\mu}B=A(\partial^\mu B)-(\partial^\mu A)B$, $A^{[\mu} B^{\nu]}=A^\mu B^\nu-A^\nu B^\mu$, $\sigma^{\mu\nu}\gamma_5=\frac{i}{2}\varepsilon^{\mu\nu\rho\sigma}\sigma_{\rho\sigma}$.} and observable Michel-spectrum shapes for
$\mu^-\to e^-+X\bar X$, in the limit $m_e=m_X=0$, after summing over the electron spin. $S,\,V,\,T$ currents in the $\mu\to e$ part are as in eqs.~(\ref{eq:scalarops}).
The spectrum is written as\\
$
\frac{d^2\Gamma}{dx\,d\cos\theta}
=
{\cal N}_{s,a}
\left(|g_L^{(a)}|^2+|g_R^{(a)}|^2\right)
\left[
F_0^{(s,a)}(x)
+
P_\mu\cos\theta\,
\chi_a F_P^{(s,a)}(x)
\right],
\qquad
\chi_a=
\frac{|g_R^{(a)}|^2-|g_L^{(a)}|^2}
{|g_R^{(a)}|^2+|g_L^{(a)}|^2}.
$
\\All overall normalizations, including Wilson coefficients and numerical constants from the invisible phase-space tensors, are absorbed into ${\cal N}_{s,a}$. The SM case is recovered for $s=1/2$, $a=V$ and $\chi_a=-1$, with $f=\nu$, indistinguishable from $s=0$, $a=V$ and $\chi_a=-1$ (both highlighted in boldface, other degeneracies among non-SM shapes are also manifest from the table). For $s_X=1$, $V_{\mu\nu}$ denotes the gauge-invariant field strength of the massless vector (identical results are obtained replacing one $V$ by its dual). 
For $s_X=3/2$, $\Psi_{\mu\nu}\equiv\partial_\mu\Psi_\nu-\partial_\nu\Psi_\mu$ denotes the gauge-invariant Rarita--Schwinger field strength (identical results are obtained either adding $\gamma_5$ or changing one $\Psi$ by its dual in the invisible currents).
For $s_X=2$, $R_{\alpha\beta\rho\sigma}$ denotes the linearized gauge-invariant curvature; this row should be understood only as a formal extrapolation (identical results are obtained changing $R$ by its dual, or using the Weyl tensor, $C$, as long as we restrict to massless on-shell spin-two invisible particles).}
\label{tab:invisible_currents_shapes}
\end{table*}

\section*{Acknowledgments}
I thank Toni Pich for insightful comments on the manuscript. I am very grateful for the invitation to the 2024 Belle-II Physics Week, where discussions first made me think of this possible loophole in the standard interpretation. I have enjoyed the hospitality and financial support 
of IFIC and Universitat de València during the completion of this work, which was partly funded
by Conahcyt-Secihti (Mexico) by project CBF2023-2024-3226, and by MCIN/AEI/10.13039/501100011033
(Spain), grants PID2020-114473GB-I00 and PID2023-146220NB-I00, and by Generalitat Valenciana (Spain),
grant PROMETEO2021/071, project CIESGT-2024-21 and 
the plan GenT program CIDEGENT/2021/037.

\bibliographystyle{apsrev4-2}
\bibliography{main}
\end{document}